\documentclass[
 reprint,
superscriptaddress,
nofootinbib,
 amsmath,amssymb,
 aps, 
]{revtex4-1}
\usepackage{amsmath,amssymb}
\usepackage{amsthm}
\usepackage{mathtools}
\usepackage[noend]{algorithmic}
\usepackage[ruled,vlined]{algorithm2e}
\usepackage{setspace}
\usepackage{url}
\usepackage{makeidx}
\usepackage{enumerate}
\usepackage{graphicx,float,psfrag,epsfig,caption}
\usepackage[usenames,dvipsnames,svgnames,table]{xcolor}
\definecolor{darkgreen}{rgb}{0.0,0,0.9}
\usepackage[pagebackref,colorlinks=true,pdfpagemode=UseNone,citecolor=OliveGreen,linkcolor=BrickRed,urlcolor=BrickRed,pdfstartview=FitH]{hyperref}
\usepackage{epstopdf}
\usepackage{color}
\usepackage{xr}
\usepackage{subfig}
\usepackage{caption}
\usepackage{graphicx}
\usepackage[utf8]{inputenc}
\usepackage{bm}
\usepackage{cancel}
\usepackage{subcaption}
\usepackage{algorithm2e}

\usepackage{scalerel,stackengine}

\stackMath
\newcommand\reallywidehat[1]{%
\savestack{\tmpbox}{\stretchto{%
  \scaleto{%
    \scalerel*[\widthof{\ensuremath{#1}}]{\kern.1pt\mathchar"0362\kern.1pt}%
    {\rule{0ex}{\textheight}}
  }{\textheight}%
}{2.4ex}}%
\stackon[-6.9pt]{#1}{\tmpbox}%
}
\usepackage[mathscr]{euscript}

\usepackage{wrapfig}

\DeclareSymbolFont{rsfs}{U}{rsfs}{m}{n}
\DeclareSymbolFontAlphabet{\mathscrsfs}{rsfs}

\newtheoremstyle{myexample} 
    {\topsep}                    
    {\topsep}                    
    {\rm }                   
    {}                           
    {\bf }                   
    {.}                          
    {.5em}                       
    {}  

\newtheoremstyle{myremark} 
    {\topsep}                    
    {\topsep}                    
    {\rm}                        
    {}                           
    {\bf}                        
    {.}                          
    {.5em}                       
    {}  

\theoremstyle{myremark}

\theoremstyle{myremark}

\theoremstyle{myexample}

\definecolor{darkgreen}{rgb}{0.0, 0.5, 0.0}

\newcommand{\bea}{\begin{eqnarray}}
\newcommand{\eea}{\end{eqnarray}}
\newcommand{\<}{\langle}
\renewcommand{\>}{\rangle}
\newcommand{\E}{{\mathbb E}}

\def\fr{\frac}
\def\fr12{\frac{1}{2}}

\def\bS{{\boldsymbol S}}

\def\eps{{\varepsilon}}

\def\bx{{\boldsymbol{x}}}

\def\bb{{\boldsymbol{b}}}
\def\bs{{\boldsymbol{s}}}

\def\bJ{\boldsymbol{J}}

\def\de{{\rm d}}

\def\E{{\mathbb E}}

\def\<{\langle}
\def\>{\rangle}

\def\by{{\boldsymbol{y}}}

\def\b0{{\boldsymbol{0}}}

\def\br{{\boldsymbol r}}

\allowdisplaybreaks
\usepackage{graphicx}

\begin{document}

\title{Chaos in high-dimensional dynamical systems with tunable non-reciprocity}

\author{Samantha J. Fournier}
    \email{samantha.fournier@ipht.fr}
\author{Pierfrancesco Urbani}%
 
\affiliation{Université Paris-Saclay, CNRS, CEA, Institut de physique théorique, 91191, Gif-sur-Yvette, France}

\begin{abstract}
High-dimensional dynamical systems of interacting variables are ubiquitous in the study of complex systems. 
When the directed interactions are totally uncorrelated, sufficiently strong and non-linear, many of these systems exhibit a chaotic attractor characterized by a positive maximal Lyapunov exponent (MLE). On the contrary, when the interactions are completely symmetric, the dynamics takes the form of a gradient descent on a carefully defined cost function, and it exhibits slow dynamics and aging. In this work, we consider the intermediate case in which the interactions are partially symmetric, with a parameter $\alpha$ tuning the degree of non-reciprocity. 
We show that for any value of $\alpha$ for which the corresponding system has non-reciprocal interactions, the dynamics lands on a chaotic attractor. Correspondingly, the MLE is a non-monotonous function of the degree of non-reciprocity. This implies that conservative forcing deriving from the gradient field of a rough energy landscape can enhance chaotic activity.
\end{abstract}

\maketitle

\paragraph*{Introduction --}
Many complex systems, from biological neural networks to ecosystems, are described in terms of degrees of freedom–the state of the neurons or the size of the population of different species in an ecosystem–that interact in an heterogeneous way. In many cases, these systems are effectively high-dimensional: the neural network in the brain is highly connected and the vast majority of ecosystems are well mixed, meaning that species are strongly interacting with each other.
Moreover, the hallmark of these systems is that their degrees of freedom interact via non-reciprocal couplings: for example the effective influence of a given neuron on the firing rate of one of its postsynaptic neurons is not the same as for the reverse process. The same is true for ecosystems with a prey-predator structure. 
Therefore, models of complex systems with directed interactions cannot avoid taking into account non-reciprocity in the couplings between degrees of freedom.

The study of non-reciprocal high-dimensional dynamical systems was started forty years ago and pioneered by Sompolinsky, Crisanti and Sommers in the context of neural networks \cite{sompolinsky1988chaos}, see also \cite{crisanti1987dynamics,crisanti1988dynamics, crisanti2018path}. More recently, non-reciprocity in low dimensional systems has been also investigated, see for example \cite{fruchart2021non, garces2025phase}.

Consider a network of $N\gg 1$ neurons described by their membrane potentials $\{x_i\}_{i=1,\ldots,N}$ and interacting via a set of synaptic connections encoded in a matrix $\bJ$ whose elements $J_{ij}$ encode how the the firing rate of neuron $j$ changes the membrane potential of neuron $i$. 
The dynamics of the system is described by a set of non-linear ordinary differential equations (ODEs)
\begin{equation}
    \dot x_i(t) = -x_i(t) +g\sum_{j(\neq i)=1}^NJ_{ij}\,\phi(x_j(t)) \equiv F_i(t),
\label{SCS}
\end{equation}
where $g$ is a constant tuning the strength of the interactions between neurons, and $\phi$ a non-linear activation function that provides the firing rate of a neuron given its current membrane potential. We call the model in Eq.~\eqref{SCS}, the SCS model.
In \cite{sompolinsky1988chaos}, the properties of this dynamical system were studied under a crucial assumption: the elements of the matrix $\bJ$ are extracted independently from the same probability distribution. This implies that $J_{ij}$ and $J_{ji}$ are uncorrelated and therefore neurons interact in a fully non-reciprocal way.
This statistical structure allows considerable simplifications in the analysis of the dynamics in the $N\to \infty$ limit and allows to show that the stationary state of the dynamics–when $g>g_c$, being $g_c$ a critical value–is a chaotic attractor characterized by a positive maximal Lyapunov exponent (MLE) \cite{crisanti2018path}.

\begin{figure*}
    \centering
    \centering
    \includegraphics[width=0.42\textwidth]{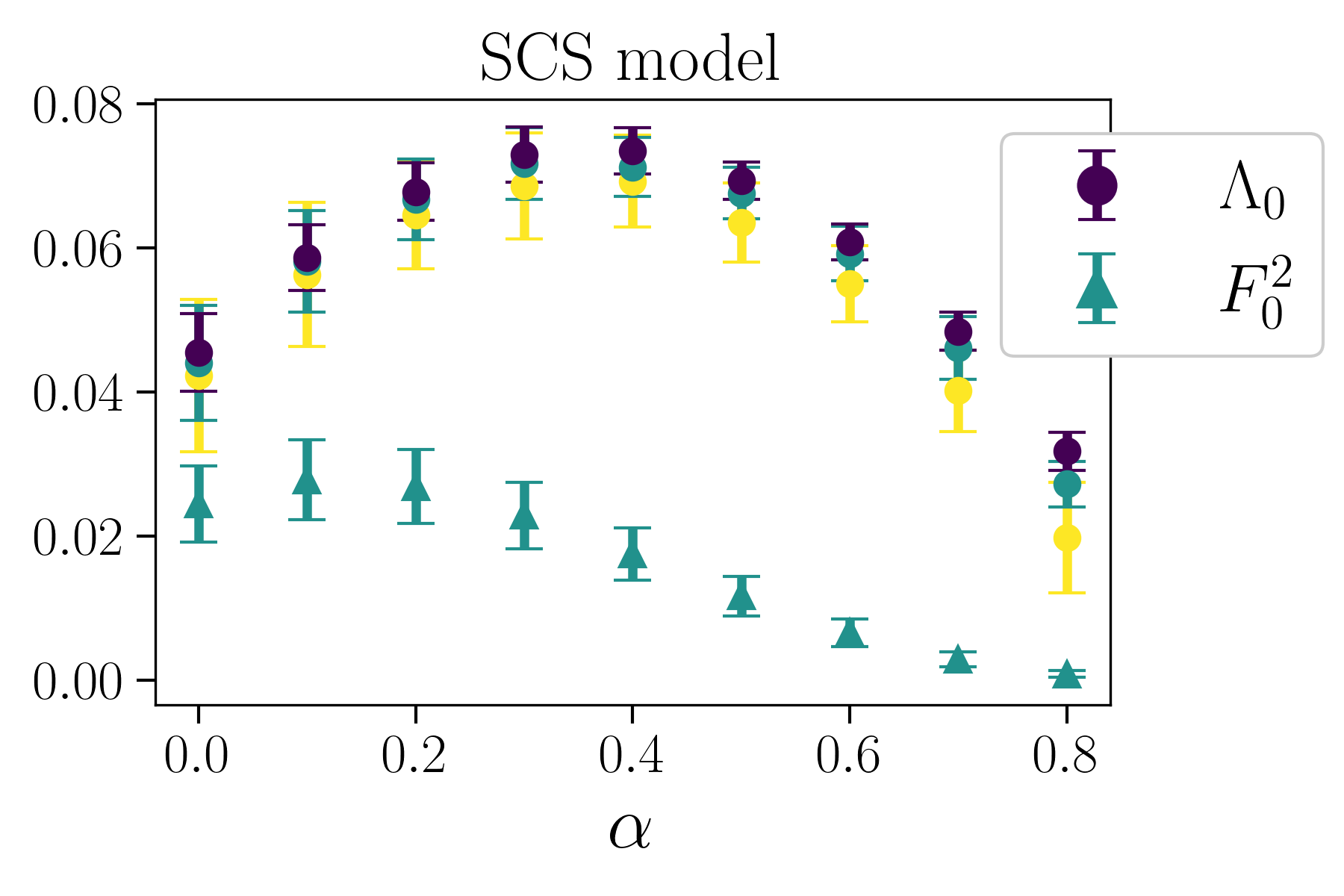}
    \hspace{1cm}
    \centering
    \includegraphics[width=0.42\textwidth]{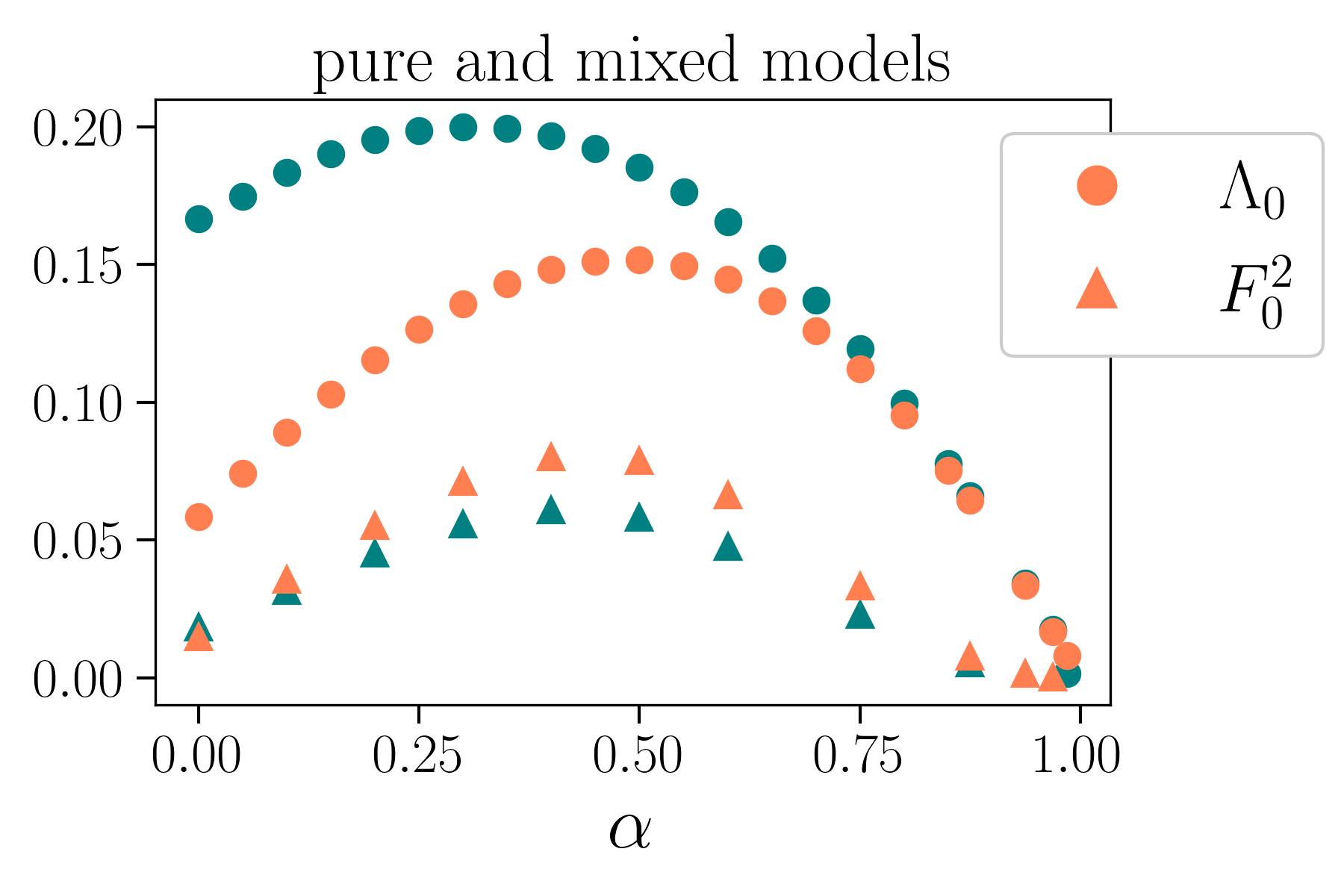}
    \caption{The MLE and the force $F_0^2$ in the stationary state. (Left) Results for the SCS model as extracted from numerical simulations. The MLE was computed using the Bennetin-Wolf algorithm \cite{fournier2025highdimensionaldynamicalsystemscoexistence} for different system sizes  $N=2500,\,5000,\,10000$ (from light to dark shades of viridis) and averaging over $N_s=100$ samples. The force $F_0^2$ was computed with $N=5000$ neurons and $N_s=1000$ samples. Parameters are: $g=1.5$, $dt=0.1$. (Right) Results for the pure and mixed models, shown in teal and coral respectively, are extracted by solving the DMFT. Parameters are: $g=0.5$ (pure model), $g=0.6$ (mixed model), $dt=0.1$.}
    \label{fig:Lyapunov summary}
\end{figure*}
However, a more realistic situation is one where $J_{ij}$ and $J_{ji}$ are correlated \cite{song2005highly}. This implies that the interactions between neurons is not completely random but has some degree of reciprocity, depending on the amount of correlations. 
It can be easily shown that, in the extreme case in which $J_{ij}=J_{ji}$ and the couplings are fully reciprocal/symmetric, the dynamics in Eq.~\eqref{SCS} admits a Lyapunov function.
Furthermore, this Lyapunov function is rough, and the system exhibits slow dynamics and aging, a phenomenology typical of spin glasses \cite{sompolinsky1981dynamic, sompolinsky1982relaxational, mezard1987spin} and which is completely different from high-dimensional chaos.
In the intermediate case in which $J_{ij}$ and $J_{ji}$ are correlated but not identical, the situation is more complicated and still unclear. 
It has been shown that, as soon as there is a small degree of non-reciprocity, the slow dynamics associated to the existence of gapless local minima of the Lyapunov function corresponding to the reciprocal part of the couplings is washed out \cite{crisanti1987dynamics,crisanti1988dynamics,fournier2025non}. However, if the dynamics is initialized in a local gapped minimum of the Lyapunov function (a large deviation initialization), the non-reciprocal part of the dynamics–as soon as its strength is not too large–induces a dynamic attractor in the basin of the local minimum \cite{cugliandolo1997glassy}. 
In both cases, the corresponding stationary state is described by a time translational invariant (TTI) dynamics but its properties are to a large extent unknown. In this work, we focus on the problem of understanding whether the dynamics of this stationary state is chaotic and we characterize in detail its properties.

Studying this question in the original SCS model is rather prohibitive from the theoretical point of view. Therefore, we consider a simple class of non-linear dynamical systems with tunable non-reciprocity that have been shown to share the same physics as the SCS model under a wide variety of settings and deformations \cite{fournier2023statistical,fournier2025highdimensionaldynamicalsystemscoexistence}. We show that (i) as soon as a degree of non-reciprocity is added to the dynamics, the corresponding stationary state is chaotic with a positive MLE, and (ii) this quantity is a non-monotonous function of the degree of non-reciprocity. 
Our results are summarized in Fig.\ref{fig:Lyapunov summary}, 
where we compare the result of our analysis on the dynamical systems that we consider to numerical simulations on the SCS model, confirming the generality of our conclusions.

\paragraph*{The model --}
We consider a high-dimensional non-linear dynamical system for a set of $i=1,\,\ldots,\,N$ degrees of freedom collected in the vector $\bx(t)$. Their dynamical evolution is described by a set of first-order ODEs
\begin{equation}
\begin{split}
    \partial_t x_i(t) &= -\mu(t)x_i(t) + g\, f_i(\bm{x}(t)) \equiv F_i(\bm{x}(t))\\
    \bx(0)&\sim {\cal N}(0,\mathbf{1}_N)\:,
    \label{dyn_sys}
\end{split}
\end{equation}
where $\bm{f}$ is a random force field whose strength is tuned by the parameter $g$. This force field is split as
\begin{equation}
    f_i(\bx(t)) = \sqrt{1-\alpha}\, r_i(\bm{x}(t))  + \sqrt{\alpha}\, s_i(\bm{x}(t)),
\end{equation}
where $\br$ encodes non-reciprocal (i.e. asymmetric) interactions between degrees of freedom, while $\bs$ represents reciprocal (i.e. symmetric) interactions. The parameter $\alpha\in[0,1)$ thus tunes the degree of reciprocity in the interactions. We consider $\alpha<1$ given that, for $\alpha=1$, Eq.~\eqref{dyn_sys} describes the relaxational dynamics in a rough energy landscape which gives rise to aging \cite{cugliandolo1993analytical,cugliandolo2023recent,parisi2020theory}. 

\begin{figure*}[ht]
    \centering
    \centering
    \includegraphics[width=0.48\textwidth]{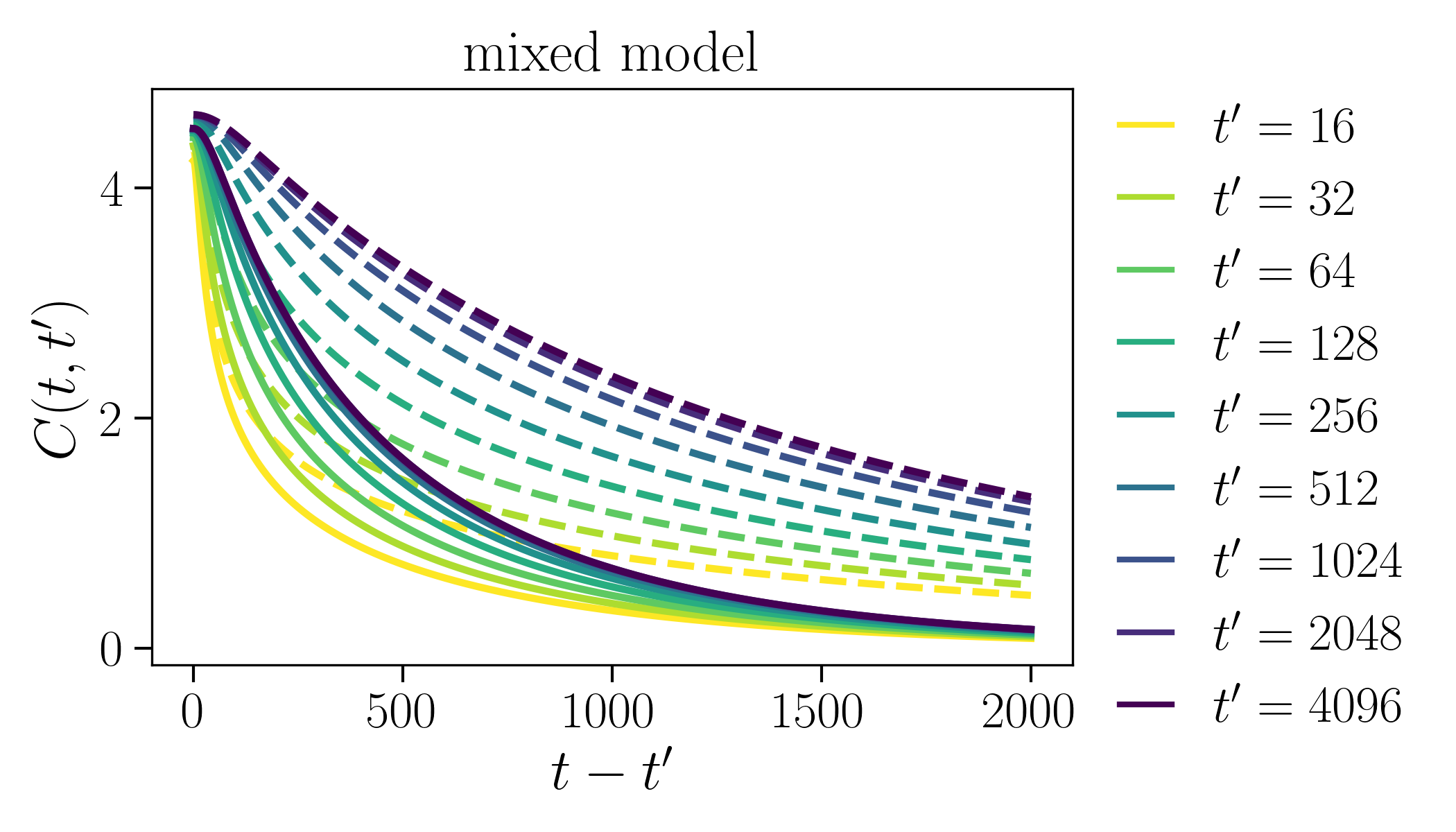}
    \hspace{0.5cm}
    \centering
    \includegraphics[width=0.48\textwidth]{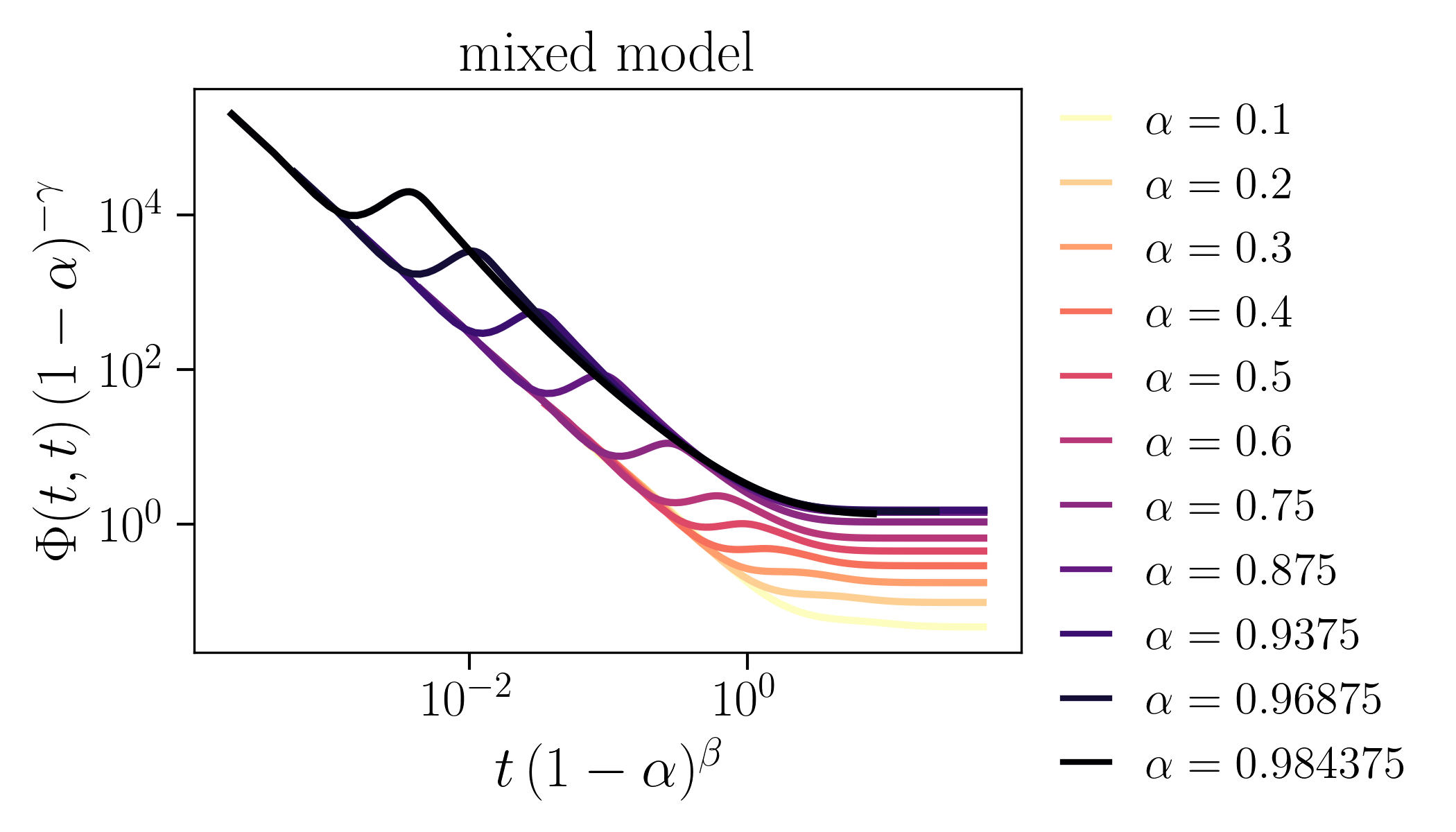}
    \caption{(Left) The correlation function $C(t,t')$ as a function of $t-t'$ and increasing $t'$. Full lines are for $\alpha=0.96875$, while dashed lines are for $\alpha=0.984375$. (Right) The norm of the driving force of the dynamical system $\Phi(t,t)$ as a function of time and for increasing $\alpha$. In both plots, results are shown for the mixed model with $g=0.6$.}
    \label{fig2}
\end{figure*}

In this work, we consider the case in which both $\br$ and $\bs$ are random Gaussian fields with a rotationally invariant covariance structure given in terms of a scalar function $h(z)$:
\begin{equation}
\begin{split}
    &\mathbb{E}[s_i(\bx)s_j(\by)]= h'\left(\frac{\bx\cdot \by}{N}\right) \frac{y_i x_j}{N}+\delta_{ij} h\left(\frac{\bx\cdot \by}{N} \right)\\
    &\mathbb{E}[r_i(\bx)r_j(\by)]=\delta_{ij}h\left(\frac{\bx\cdot \by}{N}\right)\:.
\end{split}
\end{equation}
If the function $h(z)$ is purely linear, such covariance structure corresponds to a forcing term of the form $\sum_{j}M_{ij}x_j$ with $M_{ij}=\sqrt{\alpha} M_{ij}^s+\sqrt{1-\alpha}M_{ij}^a$. Both $M_{ij}^s$ and $M_{ij}^a$ are Gaussian random variables with zero mean and unit variance. However, $M_{ij}^s=M_{ji}^s$, giving rise to reciprocal interactions, while $M_{ij}^a$ and $M_{ji}^a$ are uncorrelated, giving rise to non-reciprocal couplings.

The case of linear $h$ corresponds to an integrable dynamical system \cite{stariolo2025zero} and therefore has little to do with non-linear dynamical systems such as the SCS model. 
Therefore, we focus on the simplest non-linear model where $h(z) = g_1^2 z + 2g_2^2 z^2$ is quadratic. The specific realization of the force field in terms of random Gaussian matrices and tensors is given in the End Matter (EM).
Finally, Eq.~\eqref{dyn_sys} contains a confining term proportional to $\mu(t)$. We will assume that $\mu(t)$ is a function of $t$ only through the norm $|\bx(t)|^2/N$, which makes it rotationally invariant. In other words, $\mu(t) =\hat \mu\left(|\bx(t)|^2/N\right)$,
where $\hat\mu(.)$ is an arbitrary function whose only requirement is that it should suppress the divergence of $\bm{f}$ as $|\bx(t)|^2/N\to\infty$. For simplicity, we take $\hat\mu(z)=\mu_0+z$. We consider two instances of the model that we dub the \textit{pure} and \textit{mixed} models, defined by 
\begin{equation*}
\begin{split}
    \mathrm{pure\: model}&:\; \mu_0=0, \; g_1=0, \; g_2=1\\
    \mathrm{mixed\: model}&:\; \mu_0=1, \; g_1=2, \; g_2=1\:.
\end{split}
\end{equation*}

The model in Eq.~\eqref{dyn_sys} was studied extensively in \cite{fournier2025non}, where an analysis of the landscape of equilibria  of the dynamical system \cite{wainrib2013topological, belga2021nonlinearity, yang2025relationship, ros2022high, ben2021counting} was compared to the properties of the asymptotic attractors of the dynamics.
Furthermore, in \cite{fournier2025highdimensionaldynamicalsystemscoexistence}, the stationary dynamics of Eq.~\eqref{dyn_sys} was studied in the  $\alpha=0$ case, and it was shown that the phenomenology of the model coincides with the one of Eq.~\eqref{SCS} in a wide variety of settings and deformations. In this work, we are interested in characterizing the asymptotic attractors of the dynamics when $\alpha\in[0,1)$. In order to do this, we use dynamical mean field theory (DMFT) \cite{cugliandolo2023recent}.

\paragraph*{Dynamical Mean Field Theory --}
\label{sect: DMFT}
The dynamical system in Eq.~\eqref{dyn_sys} can be treated via DMFT using the tools developed in \cite{fournier2025highdimensionaldynamicalsystemscoexistence}. In the $N\to \infty$ limit, one can reduce the study of the full dynamical system to the one of a self-consistent stochastic process 
\begin{equation}
\label{eq: effective single site dynamics}
\begin{split}
    &\partial_t x(t) = -\mu(t) x(t) + \eta(t)\\
    &+ \alpha \int_0^t ds\, g^2 h'(C(t,s)) R(t,s) x(s) \equiv F(t),
\end{split}
\end{equation}
where the noise $\eta(t)$ is Gaussian with mean 0 and covariance
$\langle \eta(t)\eta(t')\rangle=g^2\,h(C(t,t'))$.
The correlation and response functions are defined respectively by $C(t,t') =\langle x(t)x(t')\rangle$ and $R(t,t') = \langle \delta x(t)/\delta \eta(t')\rangle$, where the brackets denote averages over the noise $\eta(t)$.
The structure of the problem in Eq.~\eqref{eq: effective single site dynamics} is self-consistent: the statistical properties of the noise have to be determined from the solution of the stochastic process itself.
However, at variance with \cite{sompolinsky1988chaos}, the models that we consider allows further simplifications since correlation and response function obey a set of closed non-linear partial differential equations given by
\begin{widetext}
\begin{equation}
\label{eq: DMFT equations}
\begin{split}
    &\partial_t C(t,t') = -\mu(t) C(t,t')  + g^2\int_0^{t'} \de s\,  h(C(t,s)) R(t',s) + \alpha \,g^2\int_0^t \de s\,  h'(C(t,s)) R(t,s) C(t',s)\equiv L_C(t,t')\\
  &\partial_t R(t,t') = -\mu(t) R(t,t') + \alpha\, g^2 \int_{t'}^t \de s\, h'(C(t,s)) R(t,s) R(s,t') + \delta(t-t') \;\;\;\;\; t\geq t'\\
  &\frac{\de C(t,t)}{\de t} = 2L_C(t,t)\:.
  \end{split}
\end{equation}
\end{widetext}
Since $\mu$ depends only on the norm of $\bx$, there is an additional constitutive equation $\mu(t)=\hat \mu(C(t,t))$.

The DMFT equations \eqref{eq: DMFT equations} can be integrated numerically with a simple Euler discretization scheme and we choose a time-step $dt=0.1$ for the presentation of our results. This choice does not affect the overall picture as far as $dt$ is sufficiently small. This discretization can be viewed either as an approximate solution or an almost exact\footnote{\parbox[t]{\columnwidth}{The discrete time DMFT equations contain an additional term of order $dt^2$ that has to be added to the equations for the propagation of $C(t,t)$, see \cite{fournier2025highdimensionaldynamicalsystemscoexistence} for details.}} DMFT treatment of the Euler-discretized dynamical system in Eq.~\eqref{dyn_sys}.

\begin{figure*}[t]
    \centering
    \includegraphics[width=1\textwidth]{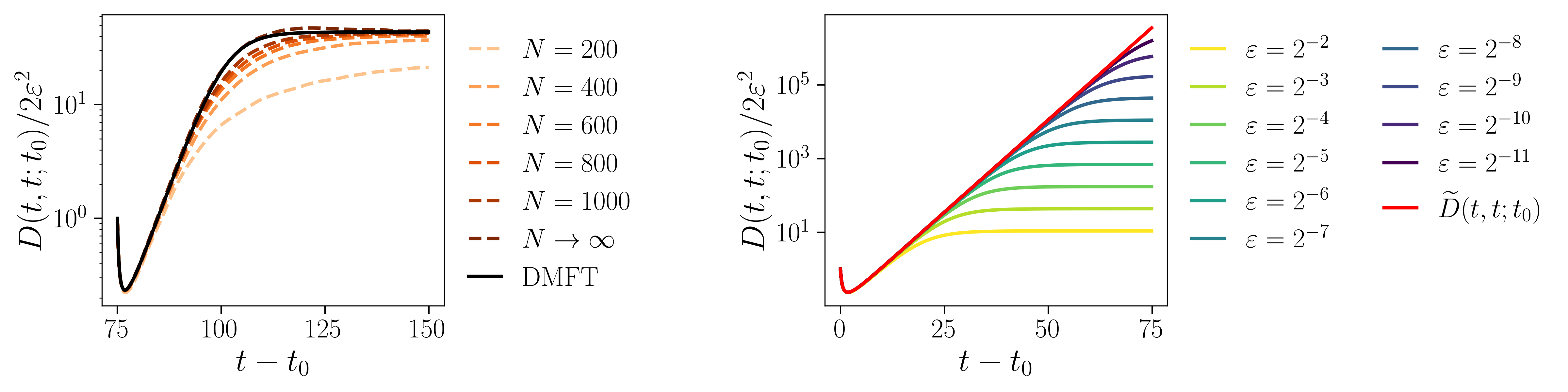}
    \caption{(Left) Comparison between numerical simulations on increasing system size (dashed colored lines) and DMFT (black line) on $D(t,t;t_0)/2\eps^2$ for $\eps=2^{-3}$. The prediction for $N\to\infty$ is obtained from a polynomial extrapolation. (Right) The behavior of $D(t,t;t_0)/2\eps^2$ for decreasing $\eps$ (viridis lines) compared to the $\eps\to 0$ limit, $\tilde D(t,t;t_0)$ (red line). In both plots, results are shown for the mixed model with $g=0.6$, $\alpha=0.2$ and $t_0=75$, $dt=0.1$.}
    \label{fig:D_and_Dtilde}
\end{figure*}

The solution of the DMFT equations gives access also to the asymptotic force driving the stationary dynamical attractor.
In particular, one can compute $\Phi(t,t')=\sum_iF_i(\bx(t))F_i(\bx(t'))/N$, see \cite{fournier2025non, zhang2025kinetic}. The precise expression of $\Phi$ is given in the EM.

In Fig.~\ref{fig2}, we plot the result of the numerical integration of the DMFT equations for increasing values of $\alpha$. For small $\alpha$, the system lands on a TTI attractor where $C(t,t')$ becomes a function of $t-t'$ (Fig.~\ref{fig2}-left). However, the relaxation time to the stationary state increases when $\alpha\to 1$. In order to characterize this properly, we plot $\Phi(t,t)$ as a function of $t$ for different values of $\alpha$. At any $\alpha<1$, the stationary state is characterized by $\lim_{t\to \infty}\Phi(t,t)=F^2_0>0$. Furthermore, we find that for $t\gg 1$ we have 
\begin{equation}
    \Phi(t,t)=(1-\alpha)^\gamma \varphi(t(1-\alpha)^\beta)
    \label{scaling_force}
\end{equation}
with $\gamma=5/2$ and $\beta=3/2$. The exponent $\beta$ coincides with the one found in \cite{BerthierKurchanTwoTimescales} on a related model. The emergence of the scaling form in Eq.~\eqref{scaling_force} is shown in Fig.\ref{fig2}-right for the mixed model, and is the same for the SCS and pure models (Fig.~\ref{fig: force scaling plots SCS and pure models} in the EM).

\paragraph*{Maximum Lyapunov exponent --}
\label{sect: MLE}
For $\alpha=0$, the dynamical system is chaotic and one can characterize the corresponding MLE exactly \cite{fournier2025highdimensionaldynamicalsystemscoexistence}. We investigate whether this remains true also at $\alpha>0$ by computing the MLE in the stationary state.
The MLE is a measure of the system's sensitivity to perturbations. Consider therefore two copies $\bx^a(t)$ ($a=1,2$) of the dynamical system in Eq.~\eqref{dyn_sys}. These two copies are evolved starting from the same initial condition at $t=0$ and thus have exactly the same trajectory up to time $t_0$, where they each get slightly perturbed independently from one another. This perturbation reads $\bx^a (t_0^+) = \bx(t_0)+ \varepsilon\, \bb^a$,
where $\bb^a$ is a random vector which, since we will consider infinitesimal perturbations ($\varepsilon\to0$), we can take as a zero-mean, unit-variance Gaussian vector without loss of generality. If the system is chaotic, the distance $\delta \bx(t) = \bx^1(t) - \bx^2(t)$ between the two copies will grow exponentially at long times after the perturbation, $\delta \bx(t) \simeq \eps\, (\bb_1-\bb_2)\, e^{\Lambda_0 (t-t_0)}$ for $t\gg t_0$.
The rate of the exponential growth $\Lambda_0$ is the MLE. Within DMFT, we can derive a flow equation for the squared distance in the limit of infinitesimal perturbations
\begin{gather*}
    \tilde{D}(t,t;t_0)= \lim_{\eps\to 0} \frac1{2\eps^2}\lim_{N\to \infty} \frac 1N \sum_{i=1}^N {(\delta x_i(t))^2}\:.
\end{gather*} This is done by writing the path-integral formulation of the twice-replicated system and using the statistical symmetry of the two copies $\bx^a(t)$ ($a=1,2$). The large-$N$ dynamics of the twice-replicated system is thus effectively described by two self-consistent stochastic processes which read
\begin{equation}
\label{eq: effective process for twice replicated system}
\begin{split}
    &\partial_t x^a(t) = -\mu(t) x^a(t) + \varepsilon\, b^a \delta(t-t_0)\\
    &+ \alpha\, g^2\int_0^t \de s\,  h'(C^{aa}(t,s)) R_d(t,s) x^a(s)+ \eta^a(t),
\end{split}
\end{equation}
where permutation symmetry in the two replicas implies that the correlation function has a simple structure $    C^{ab}(t,t') = \delta_{ab}\, C_d(t,t') + (1-\delta_{ab}) C_o(t,t')$.
The equations for $C_d(t,t')$, $C_o(t,t')$ and $R_d(t,t')$ are shown in the EM and can be integrated numerically to extract $D(t,t;t_0)=C_d(t,t)-C_o(t,t)= |\delta \bx(t)|^2/(2N)$. In figure \ref{fig:D_and_Dtilde}-left, we plot the behavior of $D(t,t_0)$ at $\alpha$ fixed and in the stationary state, and we compare the DMFT integration with numerical simulations at finite and increasing $N$, showing a good convergence. We see that $D(t,t;t_0)$ has a first transient regime, followed by a exponential growth that saturates on an $\varepsilon$-dependent plateau. In the $\varepsilon\to 0$ limit, the dynamics has an indefinite exponential growth and this is shown in Fig.\ref{fig:D_and_Dtilde}-right. The corresponding flow equation for $\tilde D$ is
\begin{equation}
\begin{split}
    &\partial_t \tilde{D}(t,t';t_0) = -\mu(t) \tilde{D}(t,t';t_0)  \\
    & +g^2\int_{t_0}^{t'} \de s\, h'(C(t,s)) \tilde{D}(t,s;t_0) R_d(t',s) \\
    &+ \alpha\, g^2 \int_{t_0}^t \de s\, h'(C(t,s)) \tilde{D}(t',s;t_0) R_d(t,s)
\end{split}
\end{equation}
from which we get the MLE as
\begin{equation}
    \Lambda_0 = \lim_{t-t_0\to \infty}\frac 1{2(t-t_0)} \ln \tilde D(t,t;t_0)\:.
\end{equation}
The result of this analysis is shown in Fig.\ref{fig:Lyapunov summary}-right. We find that: (i) $\Lambda_0$ is a non-monotonous function of $\alpha$, and (ii) it qualitatively reproduces  the numerical results on the SCS model. The behavior of the MLE as a function of $\alpha$ is compared to the one of $F_0^2$, which is also non-monotonous but with a maximum at a critical $\alpha$ which does not coincide with the one for which the MLE is maximal. For the SCS model, the non-monotonous behavior of the MLE can be qualitatively interpreted by noting that the quiescent-to-chaotic transition line $g_c(\alpha)$ is a monotonously decreasing function of $\alpha$~\cite{Brunel_correlations}. This result follows from the linear stability analysis of the fixed point $\bx=\bm{0}$, which, for the mixed model considered in this work, yields $g_c(\alpha)=g_1^{-1}/(\mu_0+\alpha)$ with $g_1\neq0$. Thus, for the SCS and mixed models, the non-monotonous behavior of the MLE as a function of $\alpha$ at fixed $g$ can be qualitatively expected as resulting from a competition between, on the one hand, moving away from the quiescent-to-chaotic transition line $g_c(\alpha)$ and, on the other hand, approaching the chaotic-to-aging transition at $\alpha=1$. Instead, along a line of the phase diagram $(g,\alpha)$ such that $g=g_c(\alpha)+\mathrm{offset}$ is at a constant distance from the quiescent-to-chaotic transition line, the MLE is a monotonously decreasing function of $\alpha$ (not shown). The interpretation for the pure model with $g_1=0$ is more delicate since it requires a non-linear stability analysis.

All in all, the results presented in this work allow to quantitatively corroborate the picture drawn by the linear stability analysis.
They show that any finite degree of (extensive-rank) reciprocity in the dynamics does not destroy chaotic behavior, but can even amplify it. This is a non-trivial collective effect whose consequences on learning dynamics \cite{sussillo2009generating, fournier2023statistical} and generative capabilities \cite{fournier2025generative,yu2025neural} must be carefully investigated.

\paragraph*{Acknowledgements --}
The authors thank Nicolas Brunel for insightful discussions. PU acknowledges funding by the French government under the
France 2030 program (PhOM - Graduate School of Physics) with reference ANR-11-IDEX-0003.

\bibliography{refs.bib}

\widetext

\section*{End Matter}

\paragraph*{Microscopic definition of the model --}
Here we write explicitly the precise form of the driving force of the dynamical systems studied in the main text and corresponding to the function $h(z)$ that we study.
The force field has a symmetric interaction part and an asymmetric one.
The components of the asymmetric part of the force field read
\begin{equation}
\begin{split}
r_i(\bm{x}(t)) &= \sum_{j=1}^N g_1\,J^{(1)j}_i x_j(t)+ \sum_{j,k=1}^N g_2\,J^{(2)jk}_i x_j(t) x_k(t)\:.  
\end{split}
\end{equation}
The couplings $\bJ^{(1)}$ and $\bJ^{(2)}$ are, respectively, \emph{i.i.d.}~random vectors and \emph{i.i.d.}~symmetric random matrices with Gaussian statistics
\begin{equation*}
\begin{split}
    &\E [J_i^{(1)j}]=\E[{J_i^{(2)jk}}]=0,\; \E[J_i^{(1)j}J_i^{(1)k}]=\frac1N\delta_{jk},\\
    &\E[{J_i^{(2)kl} J_j^{(2)nm}}]=\frac1{N^2} \delta_{ij}(\delta_{kn}\delta_{lm} + \delta_{km}\delta_{ln})\:.
\end{split}
\end{equation*}
Due to the lack of symmetry in $\bJ^{(1)}$ and in the upper indices of $\bJ^{(2)}$, this part of the force field is non-conservative in the sense that it cannot be written as the gradient of a carefully defined potential function. It will therefore act as an out-of-equilibrium drive to the dynamics. 

On the other hand, the symmetric part of the force field is defined as
\begin{equation}
\begin{split}
s_i(\bm{x}(t)) &= \sum_{j=1}^N g_1\,S_{i}^{(1)j} x_j(t)+ \sum_{j,k=1}^N g_2\,S_{i}^{(2)jk} x_j(t) x_k(t)\:.
\end{split}
\end{equation}
The couplings $\bS^{(1)}$, $\bS^{(2)}$ are random full-rank symmetric tensors. Up to symmetries, they have independent Gaussian entries with zero mean and variance
\begin{equation*}
\begin{split}
    &\E [S_{ik}^{(1)} S_{jl}^{(1)}] = \frac{1}{N} [\delta_{ij}\delta_{kl} + \delta_{kj}\delta_{il}],\\
    &\E[{S_{ikl}^{(2)} S_{jnm}^{(2)}}]= \frac{1}{N^2}[\delta_{ij}(\delta_{kn}\delta_{lm} + \delta_{km}\delta_{ln})+ \delta_{in}(\delta_{kj}\delta_{lm} + \delta_{km}\delta_{lj}) + \delta_{im}(\delta_{kj}\delta_{ln} + \delta_{kn}\delta_{lj}) ]\:.
\end{split}
\end{equation*}

\begin{figure*}[h]
    \centering
    \includegraphics[width=0.48\textwidth]{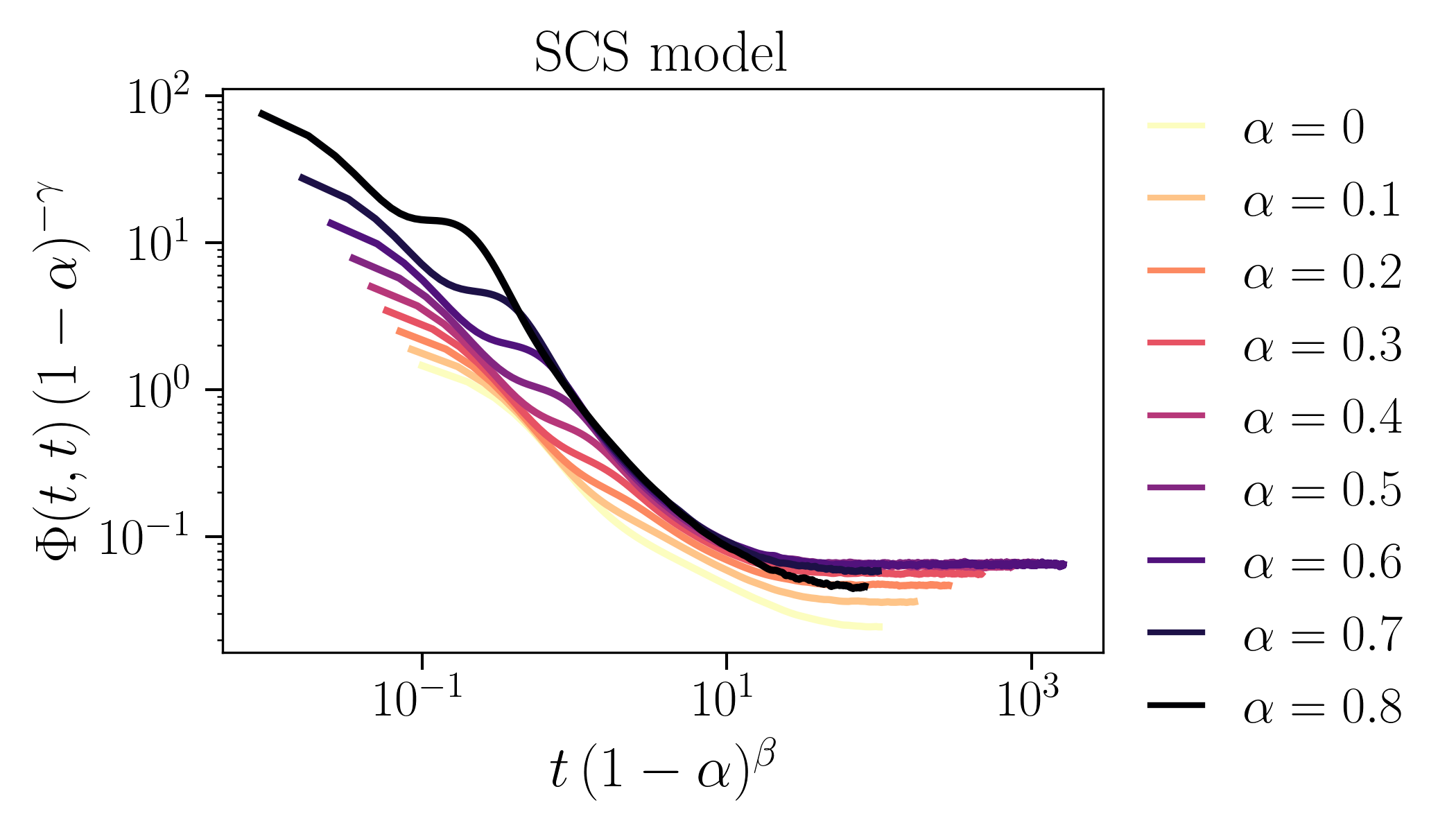}
    \hspace{0.5cm}
    \centering
    \includegraphics[width=0.48\textwidth]{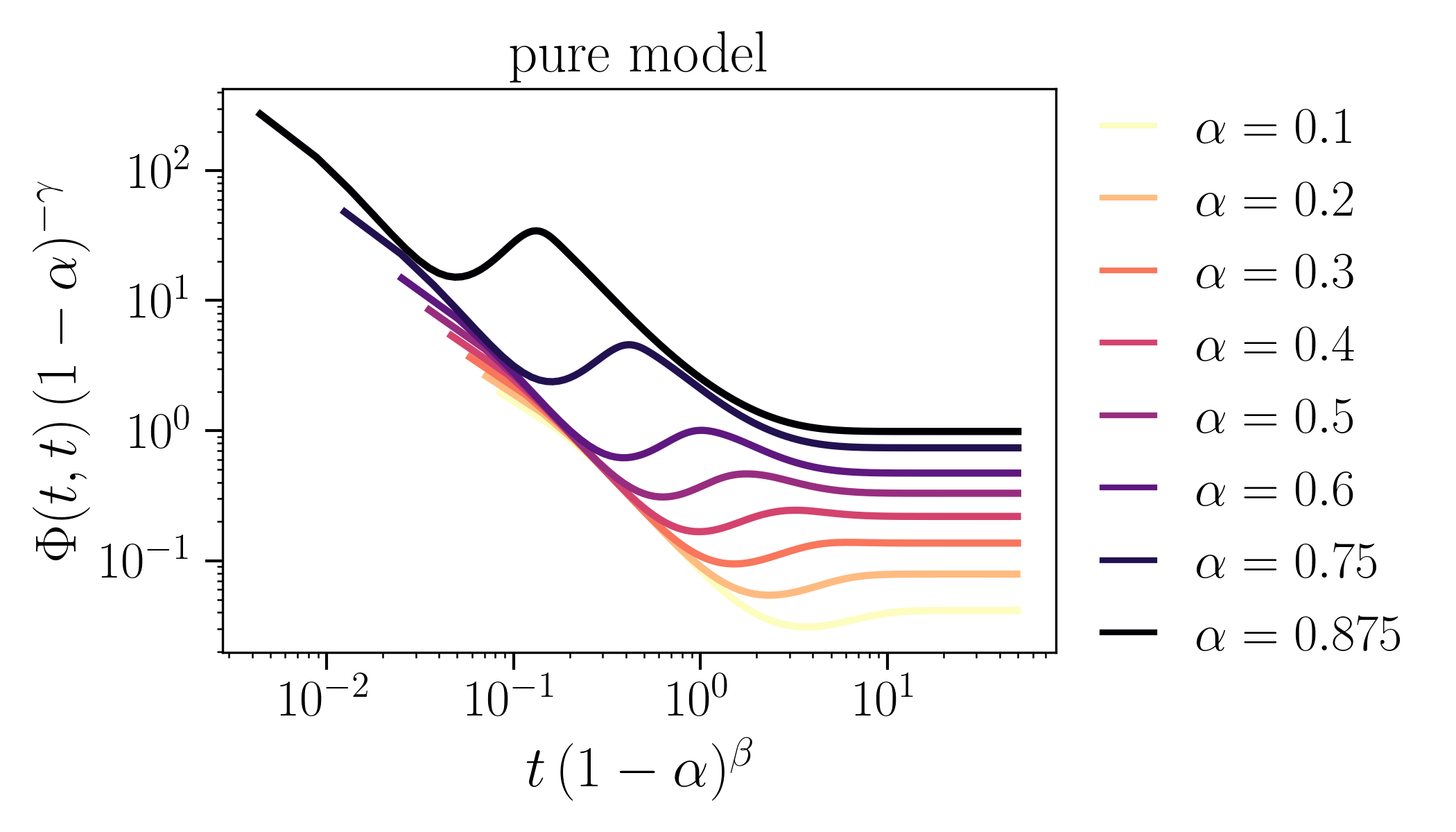}
    \caption{Scaling form of the total force driving the dynamical systems $\Phi(t,t)=\sum_i F_i^2(t)/N$ for the SCS model (Left) and the pure model (Right). Results for the SCS model were obtained by numerical simulations of the dynamics in Eq.~\eqref{SCS} with $g=1.5$, $dt=0.1$, $N=5000$ and averaging over $N_s=1000$ samples with different realizations of the disorder. Note that numerical simulations are subject to finite-size effects, particularly for large values of $\alpha$. Results for the pure model were obtained by numerically integrating the DMFT Eqs.~\eqref{eq: DMFT equations} with $g=0.5$ and $dt=0.1$. The exponents $\gamma=5/2$ and $\beta=3/2$ are the same as in Fig.~\ref{fig2}.}
    \label{fig: force scaling plots SCS and pure models}
\end{figure*}

\paragraph*{Approach to the stationary state for $\alpha\to 1$ --}
In all the models considered in this work (i.e.~the SCS, pure and mixed models), the approach to the stationary state for $\alpha\to 1$ empirically follows the same law as in Eq.~\eqref{scaling_force}. The corresponding scaling form for the force $\Phi(t,t)= \sum_{i=1}^N F_i^2(t)/N$ was shown in Fig.~\ref{fig2}-right for the mixed model. Here, in Fig.~\ref{fig: force scaling plots SCS and pure models}, we show that the same scaling form Eq.~\eqref{scaling_force} also holds for the SCS and the pure model.

\paragraph*{DMFT for the 2-replica system --}
An essential step to compute the MLE is the evaluation of the correlation structure between two, closely initialized, replicas of the dynamical system.
In the main text, we pointed out that this new system has a DMFT description in terms of a couple of self-consistent stochastic processes, as in Eq.~\eqref{eq: effective process for twice replicated system}. As for the single replica system, one can project such equations on the correlation and response functions of the system.

Here we report these equations.
The correlation function $C_d$ and $R_d$ obey the following flow equations
\begin{equation}
    \begin{split}
    \partial_t C_d(t,t') &= -\mu(t) C_d(t,t')  + \int_0^{t'} ds\, g^2 h(C_d(t,s)) R_d(t',s) + \alpha \int_0^t ds\, g^2 h'(C_d(t,s)) R_d(t,s) C_d(t',s)\equiv L_C^{(d)}(t,t')\\
    \partial_t R_d(t,t') &= -\mu(t) R_d(t,t') + \alpha \int_{t'}^t ds\, g^2 h'(C_d(t,s)) R_d(t,s) R_d(s,t') + \delta(t-t') \;\;\;\;\; t\geq t'\\
    \frac{\de C_d(t,t)}{\de t} &= 2L_C^{(d)}(t,t)+\varepsilon^2\delta(t-t_0)\:.
  \end{split}
\end{equation}

Lastly, the DMFT equations for the propagation of the off-diagonal correlation function $C_o$ reads
\begin{align}
    \begin{split}
        \partial_t C_o(t,t') &= -\mu(t) C_o(t,t') + \alpha g^2 \int_0^t \de s\,  h'(C_d(t,s)) C_o(t',s) R(t,s)+ g^2\int_0^{t'} \de s\,  h(C_o(t,s)) R(t',s)
    \end{split}\\
    \begin{split}
        \frac 12 \frac{\de C_o(t,t)}{\de t} &=  -\mu(t) C_o(t,t') + \alpha  g^2\int_0^t \de s\, h'(C_d(t,s)) R(t,s) C_o(t',s) + g^2 \int_0^{t'} \de s\, h(C_o(t,s)) R(t',s)\:.
    \end{split}
\end{align}

\end{document}